\documentclass[letter,
               keeplastbox,   
               ]{jacow}
\usepackage{pdfpages,multirow,ragged2e} %
\makeatletter%
	\ifboolexpr{bool{xetex}}
	 {\renewcommand{\Gin@extensions}{.pdf,%
	                    .png,.jpg,.bmp,.pict,.tif,.psd,.mac,.sga,.tga,.gif,%
	                    .eps,.ps,%
	                    }}{}
\makeatother

\ifboolexpr{bool{xetex} or bool{luatex}} 
 {}                                      
 {\usepackage[utf8]{inputenc}}           

\usepackage[USenglish]{babel}

\ifboolexpr{bool{jacowbiblatex}}%
 {%
  \addbibresource{jacow-test.bib}
  \addbibresource{biblatex-examples.bib}
 }{}
\begin{document}

\title{Observation of a synchro-betatron instability\\ in Fermilab booster\thanks{This manuscript has been authored by
    Fermi Research Alliance, LLC under Contract No. DE-AC02-07CH11359
    with the U.S. Department of Energy, Office of Science, Office of
    High Energy Physics.}}

\author{M. Balcewicz\thanks{balcewic@fnal.gov}, J. Eldred, Fermi National
  Accelerator Laboratory, Batavia, IL, USA 
}
	
\maketitle

\begin{abstract}
   In preparation for PIP-II, there has been interest in running the Fermilab Booster at a higher current more indicative of the PIP-II era operation. In July 2023, an experiment was performed to study collective instabilities over the transition crossing at the Fermilab Booster. Over the transition crossing, the synchrotron tune becomes small and synchro-betatron instabilities become possible. During the experiment, an intensity threshold was observed, above which a dipole instability with losses concentrated in the tail of the bunch. These losses are consistent with the Convective Instability.
\end{abstract}
\section{Introduction}
To achieve the benchmarks set by the PIP-II project, a total of $1.2$ MW\cite{cdr2017} must be delivered to the target.  This is a significant step up from the maximum of $\sim 1.0$ MW currently achievable. In order to achieve this target power, at least $6.5\times 10^{12}$\cite{cdr2017} particles must be extracted every booster cycle while not increasing the losses within the machine. 

Complicating this, as the intensity increases the beam can encounter a threshold where losses increase sharply at transition. Beyond this threshold, the vertical centroid increases in amplitude until particle loss is observed. This instability is not purely transverse as longitudinal motion also plays a significant role in the dynamics. These thresholds and the dynamics behind them must be characterized in order determine a remedy to such beam loss.

These instabilities are consistent with the so called 'Convective Instability' theorized by Burov~\cite{convective} with significant asymmetry in the transverse centroid position along the length of each bunch, with dipole amplitude and particles losses focused at the tail of each bunch.





\section{Convective Instability}
It is tempting to neglect the contributions of longitudinal motion on transverse dynamics. However, longitudinal motion has a (mostly one way) contribution to transverse motion. Energy deviations within the bunch make particles oscillate off center and off frequency. Transverse space charge forces vary with beam intensity while wakes accumulate along the length of the bunch creating spatial inhomogeneity in transverse motion. In a synchrotron, particles oscillate around the synchronous phase undergoing successive synchrotron oscillations. These oscillations move particles in the longitudinal phase space and average out some of the transverse head-tail asymmetry. 

%


It has been previously theorized\cite{convective,thesis} that when space charge forces and wakes become sufficiently large compared to the synchrotron tune, ($\Delta Q_{sc}/Q_s \gg 1$)\cite{corehalo} a saturating instability can develop. For this instability the transverse dipole motion, the motion of the beam center around the closed orbit gains a larger amplitude near the longitudinal tail in the rear of the bunch. This is known as head-tail amplification and if it becomes large enough can cause the tail of the bunch to scrape against beam pipe and cause losses. This amplification between the head and the tail of the bunch will saturate to an equilibrium amplitude, and not necessarily cause losses for a sufficiently large aperture.  This is known as the saturating convective instability (SCI). The saturating instability is generally decohered by centroid oscillation, but can lead to significant loss if some dipole kick excites motion in the bunch centroid.

Furthermore, the internal structure of the bunch does not oscillate rigidly together. Normal modes of the internal structure of this bunch evolve and couple together making some of these modes grow exponentially. Beyond an intensity threshold coupling between these modes can facilitate an absolute convective instability (ACI) where the centroid position of the bunch exponentially grows until particle losses become significant enough to reduce the beam intensity below instability threshold. After losses, such bunches will still have large head-tail amplification factors which make them vulnerable to further particle loss. As particles are lost from the tail, the bunch length will be shortened.

In the Fermilab Booster, we are below the convective instability threshold except during transition-crossing, where the instability has been observed in prior studies. During transition the synchrotron tune becomes small and the ratio between the space charge tune shift and the synchrotron tune becomes large. A sufficiently intense beam would stay above the absolute instability threshold long enough that the growth in dipole motion would drive particle loss.














\section{Experiment}


An experiment was performed in July 2023 to better understand the convective instability at the Fermilab booster. For the experiment the Booster transverse damper stripline is used as a high bandwidth dipole transverse pickup. This makes it possible to study the transverse moments of the bunch along the length of the bunch every time they pass the device using a $4$ GHz oscilloscope. Consequently the dipole motion can be measured along the length of the bunch, including any head-tail amplification.

In the Booster there are 81 bunches per beam cycle in the booster along with leaving empty a 3 bucket long 'notch'\cite{notcher} once per revolution. These bunches may shift within their individual buckets and must be centered with respect to subsequent turns to ensure a good signal resolution.

Data was taken at two main settings and an increasing intensity until instabilities are observed. The first setting was for a beam under operational conditions. The second setting was for a beam excited by a single dipole kick several hundred turns before beam transition. This objective of this kick was to drive dipole oscillations of the centroid and make them large enough to be observed over noise. If convective motion is present, it should be possible to observe the head-tail amplification between the head and the tail of the bunch. As intensity increases, so should the head-tail amplification, making the total centroid amplitude larger in turn.

\begin{figure}[!htb]
   \centering
   \includegraphics*[width=1.0\columnwidth]{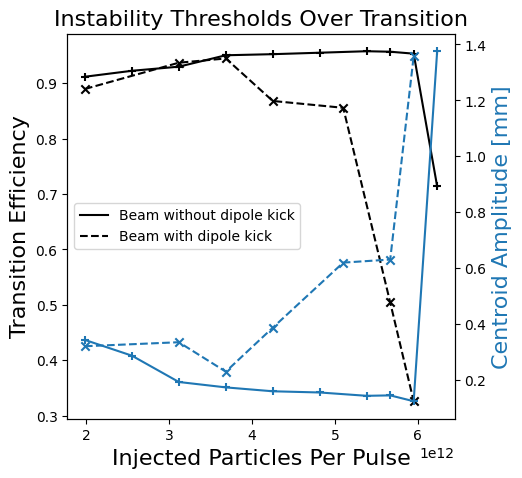}
   \caption{Booster beam transmission frequency and amplitude over transition at varying intensity. At a sufficiently high intensity ($\sim 6\times 10^{12}$ injected particle per pulse) the centroid amplitude spikes and is accompanied by significant losses. Below this threshold, only the dipole kicked beam has significant losses and centroid motion.}
   \label{threshold}
\end{figure}

Figure \ref{threshold} shows a comparison of the beam at nominal settings along with a beam experiencing a dipole kick. As intensity is increased the kicked beam centroid amplitude of the kicked beam increases, which is consistent with increasing head-tail amplification of a convective instability. For both settings there is also an intensity threshold beyond which both beams experience a spike in losses and increased centroid amplitude. This is indicative of an absolute instability which grows until particle losses push the beam below threshold and conclude the instability.



\begin{figure}[!htb]
   \centering
   \includegraphics*[width=1.0\columnwidth]{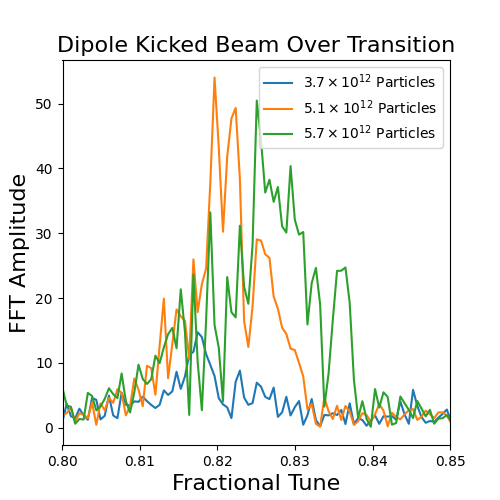}
   \caption{Vertical betatron spectrum of a kicked beam crossing transition at different intensities. Due to longitudinal motion, the spectra are comprised of several distinct sideband frequencies which shift due to collective effects and bunch distribution.}
   \label{relativefreq}
\end{figure}

\begin{figure}[!htb]
   \centering
   \includegraphics*[width=1.0\columnwidth]{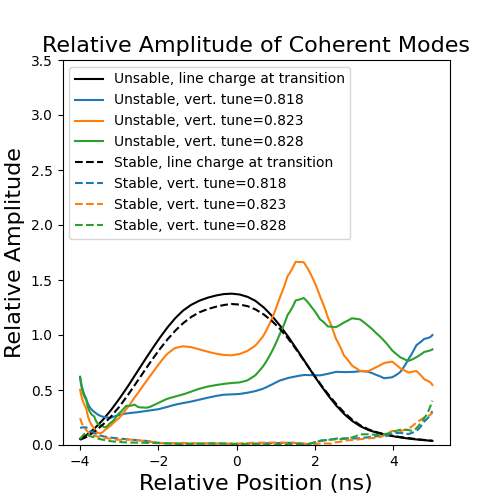}
   \caption{The vertical betatron amplitude of the three dipole modes along the length of the bunch above (solid line) and below (dashed line) the instability threshold. Past the threshold, the absolute convective instability causes the entire beam amplitude to increase beyond bunch noise. The head-tail asymmetry persists after beam losses.}
   \label{abovebelow}
\end{figure}

The dipole motion of the beam is constructed from a weighted sum of individual coherent modes, each of which vary along the length of the bunch. These sideband modes shift as the intensity of the bunch increases. Figure \ref{relativefreq} shows the dipole spectra of the kicked beam over three intensities. Without something to excite coherent motion (a kick or an instability) these betatron signals would be small and difficult to resolve. Each coherent mode has a characteristic longitudinal shape along the length of the bunch, an example of which is shown in Figure \ref{abovebelow}.


While the kicked beam demonstrates that some of these modes are in fact convective, it is important to determine whether the high intensity particle losses are due to the absolute convective instability. For a beam under operation settings without a kick, oscillations are weak and are difficult to observe over noise until the beams pass an intensity threshold (dashed lines in Figure \ref{abovebelow}). Just past the threshold, a very strong betatron oscillation frequencies can be observed, but with a notable lack of sideband structure in Figure \ref{relativefreq}. Only some modes become absolutely unstable and grow exponentially, which is consistent with our understanding of ACI. 

\section{Conclusion}


At the Fermilab booster, instabilities can be observed at approximately $6\times 10^{12}$ particles per bunch and operational settings over transition.  These instabilities are characterized by a large dipole amplitude at the longitudinal tail of the bunch which drives beam loss and are consistent with an absolutely convective instability.

While there is significant head-tail amplification, the shape and behavior of individual convective modes is still an area of active investigation. Conventionally, the amplitude of the convective instability grows toward the tail of the bunch making a cobra shape\cite{convective} (maximum transverse amplitude with a zero derivative at the bunch tail edge). For the Fermilab Booster, Figure \ref{abovebelow} shows a peak at around $2$ ns past the center of the bunch, after which the transverse amplitude decreases. This suggests that different longitudinal distributions and transverse wake impedances can change the characteristic transverse shape of the bunch while retaining the transition between saturating and absolute instabilities.



This instability drives significant losses and must be avoided during the PIP-II era to prevent beam losses.  It may be possible to fine tune chromaticity in the Booster to avoid triggering the absolute convective instability. In order to ensure the successful operation of the booster in the PIP-II era either a systematic approach must be developed to prevent the instability over transition, transition must itself must be changed. The change to a transition-less lattice or a $\gamma_t$ jump may make it possible to avoid this instability entirely.


\section{Acknowledgments}
We are grateful for the late Peter Prieto's help with the oscilloscope instrumentation. Thank you to A. Burov for his advice on better understanding the instability.  C. Bhat, S. Chaurize, and K. Triplett for their help operating the booster and setting up the damper stripline.

%
%
\ifboolexpr{bool{jacowbiblatex}}%
	{\printbibliography}%
	{%
	
    
} 

%
%


\end{document}